# Weakly collisional Landau damping and three-dimensional Bernstein-Greene-Kruskal modes: New results on old problems

C. S. Ng and A. Bhattacharjee

*Space Science Center, Institute for the Study of Earth, Oceans, and Space, University of New Hampshire, Durham, NH 03824*

F. Skiff

*Department of Physics and Astronomy, The University of Iowa, Iowa City, IA 52242*

Landau damping and Bernstein-Greene-Kruskal (BGK) modes are among the most fundamental concepts in plasma physics. While the former describes the surprising damping of linear plasma waves in a collisionless plasma, the latter describes exact undamped nonlinear solutions of the Vlasov equation. There does exist a relationship between the two: Landau damping can be described as the phase-mixing of undamped eigenmodes, the so-called Case-Van Kampen modes, which can be viewed as BGK modes in the linear limit. While these concepts have been around for a long time, unexpected new results are still being discovered. For Landau damping, we show that the textbook picture of phase-mixing is altered profoundly in the presence of collision. In particular, the continuous spectrum of Case-Van Kampen modes is eliminated and replaced by a discrete spectrum, even in the limit of zero collision. Furthermore, we show that these discrete eigenmodes form a complete set of solutions. Landau-damped solutions are then recovered as true eigenmodes (which they are not in the collisionless theory). For BGK modes, our interest is motivated by recent discoveries of electrostatic solitary waves in magnetospheric plasmas. While one-dimensional BGK theory is quite mature, there appear to be no exact three-dimensional solutions in the literature (except for the limiting case when the magnetic field is sufficiently strong so that one can apply the guiding-center approximation). We show, in fact, that two- and three-dimensional solutions that depend only on energy do not exist. However, if solutions depend on both energy and angular momentum, we can construct exact three-dimensional solutions for the unmagnetized case, and two-dimensional solutions for the case with a finite magnetic field. The latter are shown to be exact, fully electromagnetic solutions of the steady-state Vlasov-Poisson-Ampère system.

## I. Introduction

High-temperature plasmas are abundant in our universe in many astrophysical settings, as well as in man-made plasma devices in controlled fusion research. The physics of high-temperature plasmas is rich and often complicated. As is well known, one striking property that seemingly can help reduce the complexity of the problem is that a high- temperature plasma is almost collisionless (e.g., Ref. 1), since the collision frequencies within and between species are all inversely proportional to the cube of the characteristic thermal speed. A drastic but very useful approximation is to neglect collisions altogether and drop the collisional term in the right hand side of the Boltzmann equation

$$\frac{\partial f_s}{\partial t} + \mathbf{v} \cdot \frac{\partial f_s}{\partial \mathbf{r}} + \frac{q_s}{m_s}\left(\mathbf{E} + \frac{\mathbf{v}}{c} \times \mathbf{B}\right) \cdot \frac{\partial f_s}{\partial \mathbf{v}} = \left(\frac{\delta f_s}{\delta t}\right)_c , \qquad (1)$$

for the distribution function $f_s$ of all species $s$. The collisionless Boltzmann equation, or the Vlasov equation, is simply a continuity equation in the six-dimensional configuration and velocity space for $f_s$, without source or sink. The **E** and **B** in Eq. (1) are the electric and magnetic fields, which can be either external or self-generated through the Maxwell equations, or, for the purely electrostatic case with $\mathbf{E} = -\nabla \psi$, through the Poisson equation,

$$\nabla^2 \psi = -4\pi \sum_s q_s \int d\mathbf{v} f_s . \qquad (2)$$

Studies of the collisionless Vlasov equation have produced many important insights in kinetic theory, not only in plasma physics, but also in many other physical systems with similar underlying physics, such as beams in particle accelerators, galactic dynamics, ocean waves, and trapped electron devices (with applications to two-dimensional vortex dynamics), to name only a few. Two important examples are the Landau damping of linear plasma waves[3] and the existence of exact self-consistent steady-state nonlinear solutions of the Vlasov-Poisson system of equations, known as the Bernstein-Greene-Kruskal (BGK) modes[4]. There does exist a relationship between the two: Landau damping can be described as the phase-mixing of undamped eigenmodes, the so-called Case-Van Kampen modes[5,6] which can be viewed as BGK modes in the linear limit[4].

Although these concepts have been around for a long time, new extensions and unexpected new results are still being discovered over the years, too numerous to be cited exhaustively here. In this paper, we will focus on two recent results that extend our understanding of these two fundamental concepts. The first pertains to Landau damping in weakly collisional plasmas, the second to three-dimensional (3D) BGK modes.

In Section II, we will review our results on the effect of weak collision on Landau damping. It is often assumed that the presence of weak collisions will produce small and regular perturbative corrections to the collisionless results. In recent years, it has become clear that if one describes collisional processes by rigorous collision operators of the Fokker-Planck type[7,8], then weak collisions are a *singular* perturbation on the collisionless problem, and that weak collisions alter profoundly the results of the standard collisionless theory[9-11]. These theoretical results have important implications for experiments, and were in fact, motivated by laboratory experiments, which used laser-induced florescence to measure ion distribution functions with high accuracy[12].

In Section III, we will extend the usual BGK mode in one dimension (1D) to three dimensions (3D). Since the classical work of Bernstein, Greene and Kruskal[4], many papers have been written on this subject. However, the vast majority of these papers deal with 1D BGK theory. (See Ref. 13 for a recent review.). The method of constructing a 1D BGK solution is straightforward, either by specifying the distribution function or the electrostatic potential and then determining the other, as discussed in textbooks (e.g., Refs. 14, 15).

Surprisingly, exact 3D BGK solutions do not appear to have been considered until recently. Part of the reason may be because 1D BGK solutions have long been shown to be unstable in higher dimensions[16,17]. However, recently there has been renewed interest in 3D BGK solutions. This is mainly due to 3D features of solitary wave structures in space-based observations that cannot be explained by 1D BGK wave theory[18-20]. There are also results from numerical simulations suggesting the possibility of higher-dimensional BGK solutions[21-23].

3D BGK solutions have been constructed under the assumption of a strong magnetic field so that one can use some form of gyro-kinetic or drift-kinetic equation[24-27]. Solutions have also been obtained in the presence of an infinitely strong background magnetic field[28-30]. In this case, charged particles are constrained to move along magnetic field lines, and thus the 3D problem is effectively reduced to a 1D problem. Efforts on relaxing this assumption to finite-strength magnetic fields have not produced exact results so far. It has been argued that a 3D BGK solution for zero magnetic field does not exist[28]. In our recent work[31], we have shown that this last conclusion is true only if the distribution function $f$ is assumed to depend only on energy. However, if we allow $f$ to have dependence on additional constants such as the angular momentum, we can actually construct 3D BGK solutions. We will review the construction of 3D BGK solutions for an unmagnetized plasma in Section III. We will then discuss briefly in Section IV how two-dimensional (2D) BGK solutions with finite magnetic field can be constructed.

## II. Weakly Collisional Landau Damping

Landau damping of plasma oscillations in a collisionless plasma is one of the most fundamental and widely used concepts in plasma physics. Its existence has long been confirmed by experiments[32-33]. Although Landau's classic paper[3] is the standard point of departure for discussions of kinetic stability theory in most textbooks, it raises some vexing physical questions. Why should plasma oscillations damp in a collisionless plasma in which the underlying dynamics is time-reversible and non-dissipative? Besides, the damped solutions are not true eigenmodes of the system. There can be true eigenmodes for *unstable* solutions of the form $\exp(-i\omega t)$ with $\mathrm{Im}\,\omega > 0$ when the background distribution function is non-monotonic. However, why does the theory not support stable eigenmodes with $\mathrm{Im}\,\omega < 0$ when the distribution function is monotonic?

These questions have been answered within the framework of the collisionless theory[5,6,34]. A key to understanding Landau damping in collisionless plasmas is phase-mixing, made possible by the presence of a continuous spectrum, which lies on the real axis ($\mathrm{Im}\,\omega = 0$) of the complex-$\omega$ plane. This continuous spectrum is associated with a complete but singular set of eigenmodes, known as the Case-Van Kampen modes (discussed in textbooks, e.g. Refs. 2,35,36). It takes very special, that is, singular initial conditions to excite isolated Case-Van Kampen modes. In most situations of physical interest where the initial conditions are smooth, a broad and continuous spectrum of Case-Van Kampen modes is excited. The Landau-damped waves are not eigenmodes but are remnants, in the long-time limit, after a continuous and complete set of singular eigenmodes, each one of which is purely oscillatory, have interfered destructively (in the sense of the Riemann-Lebesgue theorem). In the mathematical sense, this problem is totally solved, since the set of eigenfunctions, with known analytic form, is proved to be complete.

How is this widely accepted physical picture of Landau damping modified if collisions are introduced? Ref. 8 considered the problem using an operator of the Fokker-Planck type[7]. They obtained an exact analytic solution with a dispersion relation that has a root that formally reduces to a Landau root in the limit of zero collisions. However, they did not discuss the nature of the spectrum or address the issue of completeness of the eigenmodes. Since then, many authors have considered this problem using different approaches[37-39,10]. However, a complete set of eigenfunctions has been obtained only recently[9,11].

Our studies were motivated by the results of recent experiments[12] involving a weakly collisional stable plasma. Figure 1, taken from Ref. 12, shows the linear ion distribution function perturbation. It was found that the response function is consistent with a superposition of discrete kinetic eigenmodes. Moreover, these modes appear to have damping rates smaller than expected from classical theory[37]. To gain understanding of these experimental results, we have developed a theory, which attempts to do for collisional plasmas what the Case-Van Kampen theory does for collisionless plasmas. This theory has potentially important consequences for plasma echoes[40-41] and the effect of weak collisions on such phenomena[37,38,42,43,10]. The theory essentially obtains a *complete* set of eigenfunctions, with known analytic form and a set of discrete eigenvalues. We demonstrate that the Case-Van Kampen continuous spectrum is completely eliminated, and replaced by a discrete spectrum. A subset of the eigenvalues that belongs to the discrete spectrum tends to the Landau roots as the collision frequency $v \to 0$, but the complete set of eigenvalues is much larger. Thus, the Landau roots, which are not eigenmodes of the collisionless plasma, emerge as eigenmodes of the weakly collisional plasma in the limit of zero collision. Since a real plasma must have some level of collision, no matter how small, the newly found complete set of eigenmodes, are qualitatively relevant in virtually all situations of physical interest.

Let us now look at the properties of the new set of eigenvalues and eigenfunctions in more detail. Consider, for simplicity, the electrostatic oscillations of electrons only with a uniform background of immobile ions in 1D. The linearized Boltzmann equation for the first order electron distribution function $f(x,t,v)$, with respect to a background Maxwellian distribution $f_0$ is

$$\frac{\partial f}{\partial t} + v\frac{\partial f}{\partial x} - \frac{e}{m}\frac{\partial f_0}{\partial v}E = \nu\frac{\partial}{\partial v}\left(vf + v_0^2\frac{\partial f}{\partial v}\right), \qquad (3)$$

where $\nu$ is the collision frequency, assumed to be a constant for simplicity, and $-e$, $m$ and $v_0$ are, respectively, the charge, mass, and thermal speed of the electron. Note that (3) must be solved along with the Poisson equation (2), which now takes the form

$$\frac{\partial E}{\partial x} = -4\pi e \int_{-\infty}^{\infty} dv f(x,t,v). \qquad (4)$$

Note also that we assume the linearization condition,

$$\frac{\partial f_0}{\partial v} \gg \frac{\partial f}{\partial v}. \qquad (5)$$

The right-hand-side of (3) is the linearized collision operator used in Ref. 8. Unlike the full Fokker-Planck collision term representing electron-electron as well as electron-ion collisions[44], the form in (3) neglects electron-ion collisions and assumes, furthermore, that the electron-electron collision frequency $\nu$ does not fall off with increasing velocity. However, this form does conserve particles, satisfies the H-theorem, and preserves the diffusive character of the full Fokker-Planck operator in velocity space.

Without going into the details of the derivation (which can be found in Ref. 11) here, we summarize our results. We have shown that in an initial value problem for a given spatial Fourier mode with wavenumber $k$, with suitable boundary conditions in the velocity space, a general solution can be expended by a complete set of eigenfunctions in a discrete series,

$$g(u,t) = \sum_n c_n g_n(u) \exp(-i\Omega_n t) \Theta(t), \tag{6}$$

where $\Theta(x)$ is the Heaviside step function, $g \equiv \sqrt{2} v_0 f / n_0$ is the normalized first order distribution function, $u \equiv v/(\sqrt{2} v_0)$. Here $g_n(u)$ is an eigenfunction with its Fourier transform in the velocity space $\tilde{g}_n(w) = \int_{-\infty}^{\infty} g_n(u) e^{iwu} du / \sqrt{2\pi}$ given by

$$\tilde{g}_n(w) \equiv 1 - D(w, \Omega_n) = -\alpha \left[ 1 + \frac{i\Omega_n}{\mu} d\left( \frac{1}{2\mu^2} - \frac{i\Omega_n}{\mu}, \frac{1+\mu w}{2\mu^2} \right) \right] e^{-w^2/4}, \tag{7}$$

and the corresponding eigenvalue $\Omega_n$ satisfies the dispersion relation

$$D(\Omega_n) \equiv D(0, \Omega_n) = 1 + \alpha \left[ 1 + \frac{i\Omega_n}{\mu} d\left( \frac{1}{2\mu^2} - \frac{i\Omega_n}{\mu}, \frac{1}{2\mu^2} \right) \right] = 0, \tag{8}$$

where $\mu \equiv \nu/(\sqrt{2} k v_0)$, $\alpha \equiv \omega_p^2/(k^2 v_0^2) = 4\pi n_0 e^2/(mk^2 v_0^2)$, $d(a,x) \equiv x^{-a} e^x \gamma(a,x)$, and $\gamma(a,x) \equiv \int_0^x e^{-t} t^{a-1} dt$ is the incomplete gamma function (e.g., Ref. 45). It can be shown analytically that as the collision frequency tends to zero (that is, $\mu \to 0$), a subset of these eigenvalues satisfying (8) tend to the Landau damping roots determined by the well-known relation $1 + \alpha [1 + \Omega_n Z(\Omega_n)] = 0$, where $Z$ is the plasma dispersion function (e.g., Ref. 13). However, there are additional roots which tend to values given by $\Omega_n = -i[n\mu + 1/(2\mu)]$, where $n$ is a non-negative integer, due to the presence of collisions. As claimed earlier, we thus see that Landau damped roots in a weakly collisional plasma, no matter how small the collision, corresponds to true eigenmodes. This is unlike the collisionless picture in which the Landau damping is the result of phase-mixing of singular Case-Van Kampen modes of the form

$$g_\Omega(u) = P\left[\frac{\eta(u)}{u-\Omega}\right] + \delta(u-\Omega)\left[1 - P\int_{-\infty}^{\infty}\frac{\eta(u')}{u'-\Omega}du'\right], \qquad (9)$$

for any real $\Omega$, with $P$ indicating the principal part. Note also that the coefficients $c_n$ in Eq. (6) can be determined completely in closed form. For similar results on the boundary-value problem, the reader is referred to Ref. 9.

It is useful to plot the set of eigenvalues and the shape of an eigenfunction in the new theory. Although (8) is the exact dispersion relation, calculating an eigenvalue still requires further numerical work, especially the case with small collisions, i. e., $\mu \to 0$, because the eigenfunction is increasingly singular in this limit. In Ref. 9, we have developed an efficient algorithm to do such calculations. Fig. 2, taken from Ref. 9 shows the distribution of eigenvalues for the boundary-value problem in the complex $k$-plane, for different level of $\upsilon$. We see that while a subset of these eigenvalues do tend to the Landau damping roots, there are other roots in the proximity of the imaginary axis, numerically given by $\kappa^2 = 2(i\mu - n\mu^2)$, where $\kappa \equiv \sqrt{2}kv_0/\omega$ and $n$ is a non-negative integer. Note also that the set of eigenvalues for the collisionless problem is simply the real axis, which is very different from the set of eigenvalues for the collisional problem, even in the limit of $\mu \to 0$. This is a good example of singular perturbation theory in which the nature of a problem is fundamentally changed between the case in which a small parameter ($\mu$ in our case) is infinitesimally small and the case in which it is set exactly to zero.

There is also fundamental change in the eigenfunctions. In Figure 3, we plot the eigenfunction (solid line) of the least-damped mode with eigenvalue $\Omega_0$, which tends to the Landau damping rate of about (2.545, -0.055) of the initial-value problem with $\alpha = 9$, for two values of $\mu$. We compare the numerical eigenfunction to the analytical exterior region solution $\eta(u)/(u-\Omega_0)$, also formally given by the first term in (9) for a Case-Van Kampen mode. We see that the numerical and analytical solutions agree very well in the exterior region of $u$ centered around $\Omega_{0r}$. Figure 3 might suggest that the boundary layer, where the eigenfunction departs from the exterior region solution shrinks as $\mu$ decreases, but this is not what actually occurs. In fact, it can be shown that as $\mu \to 0$, the boundary layer tends to a fixed width (of the order of $|\Omega_{0i}|$) while the eigenfunction within the layer becomes increasingly singular, with an oscillatory structure characterized by a spatial width proportional to $\mu^{-1/4}$.

Note also that the eigenfunction shown in Fig. 3 is for the least damped mode, and that it has very sharp gradients in velocity space for small $\mu$, and increasingly so as $\mu \to 0$, the collisionless limit. Initially, such sharp gradients do not show up in the distribution function since the superposition of all the modes smooth them out. However, the distribution function will eventually develop very fine velocity scales at long times, after many modes are damped away. This is also true for the thermally excited modes inside a plasma in thermal equilibrium. Thus we conclude that in the weakly collisional case for most high-temperature plasmas, the linearization condition (5) is deeply suspect even in cases where the small amplitude of waves may appear to justify linearization. Nonlinear effects can be very important in general. In the next section, we turn to another important example, BGK modes of the Vlasov equation, where nonlinear physics is essential.

**III. 3D BGK Modes**

BGK modes have been observed experimentally in both magnetized plasmas[46,47] and trapped pure electron plasmas[48]. There are also several reports on space-based observations of solitary waves or phase space holes structures in the magnetosphere and the solar wind[49-55,18-20] that could be BGK modes[56-58], including a width-amplitude relation that is more consistent with BGK modes than solitons. Numerical simulations have also shown that 1D BGK solutions that appear to be stable can be formed dynamically via a two-stream instability or nonlinear Landau damping[59-65]. Despite the success of the theory of 1D BGK modes in explaining observational data, some recent observations do show 3D features in such electrostatic structures[18-20]. For example, in Ref. 18, electrostatic solitary waves observed in the auroral ionosphere were found to have electric field components $E_\perp$ perpendicular to the background magnetic field with magnitude comparable to the parallel component, $E_\parallel$. This is inconsistent with a 1D BGK-like potential. Also, the shape of $E_\perp$ structures in the $E_\perp$ versus time plot is unipolar, while the shape for $E_\parallel$ structures is bipolar. This is consistent with the structure of a single-humped solitary potential that travels past the spacecraft along the magnetic field[28].

These 3D features seem to be consistent with recent theories that extend the 1D BGK mode to the case with strong magnetic field[24-30,58]. The physical reason for the existence of such solutions is actually related to the 1D aspect of such a theory, since charged particles are tied to a strong magnetic field like beads to a wire, which reduces a 3D problem to a 1D problem (in the limit of an infinitely strong magnetic field). So far the extension of such a solution to finite magnetic field has eluded solution, except that it has been conjectured that a BGK mode does not exist when the magnetic field is zero. In this section, we will review our recent results[31], showing that a 2D/3D BGK mode indeed does not exist if the distribution function only depends on energy. We will then demonstrate that a BGK mode solution can be constructed if the distribution function depends on energy as well as angular momentum.

We start with the electrostatic Vlasov equation for a stationary distribution function $f_s(\mathbf{r},\mathbf{v})$, along with the Poisson equation (2). For simplicity, we assume that the much more massive ions form a uniform background with constant density $n_0$, and only solve for the electron distribution. Using the normalizations $\mathbf{v} \to v_e \mathbf{v}$, $\mathbf{r} \to \lambda \mathbf{r}$, $\psi \to 4\pi n_0 e \lambda^2 \psi$, and $f_e \to n_0 f_e / v_e^3$, where $v_e$ is the electron thermal velocity and $\lambda = v_e / \omega_{pe}$ is the electron Debye length, equations (1) and (2) become

$$\mathbf{v} \cdot \frac{\partial f}{\partial \mathbf{r}} + \nabla \psi \cdot \frac{\partial f}{\partial \mathbf{v}} = 0 , \tag{10}$$

$$\nabla^2 \psi = \int d\mathbf{v} f - 1 , \tag{11}$$

where we have dropped the subscript $e$ on the electron distribution function. It is easy to see that (10) is satisfied exactly by a distribution function of the form $f = f(w)$, where $w = v^2/2 - \psi(\mathbf{r})$ is the energy. However, in order to be a truly self-consistent solution of the Vlasov-Poisson system, a distribution function of the form $f = f(w)$ must also satisfy (11). In Ref. 31 we have presented a simple proof that no such solutions exist in 2D and 3D.

In 2D, (10) and (11) give $n \equiv \nabla^2 \psi = 2\pi \int_{-\psi}^{\infty} dw f(w) - 1$. It follows that $dn/d\psi = 2\pi f(-\psi) \geq 0$. It can be shown similarly that the condition $dn/d\psi \geq 0$ is also true in 3D. This inequality leads to the conclusion that it would be impossible to obtain a physical and localized solution that satisfies the boundary conditions $\psi \to 0$ and $n \to 0$ as $|\mathbf{r}| \to \infty$. The reasons are as follows. For such a localized solution, $\psi$ must have either a local maximum that is positive, or a local minimum that is negative at a certain point in space. However, since $n = \nabla^2 \psi$ is negative (positive) in these two cases, then according to the inequality $dn/d\psi \geq 0$, $n$ must be increasingly more negative (positive) as $\psi$ decreases (increases) from its local maximum/minimum. It would thus be impossible to satisfy the boundary conditions $\psi \to 0$ and $n \to 0$ as $|\mathbf{r}| \to \infty$ simultaneously, required for a localized solution.

We now demonstrate that it is possible to construct 3D solutions if we relax the restriction that the distribution function depends only on energy. Let us consider the case in which the electrostatic potential is spherically symmetric, that is, $\psi = \psi(r)$. The distribution function will then have the form $f = f(r, v_r, v_\perp)$, where $v_\perp^2 = v_\theta^2 + v_\phi^2$, and $(r, \theta, \phi)$ are the usual spherical coordinates. With these simplifications and coordinate transformations, (10) and (11) become[66,67]

$$v_r \frac{\partial f}{\partial r} + \left( \frac{d\psi}{dr} + \frac{v_\perp^2}{r} \right) \frac{\partial f}{\partial v_r} - \frac{v_r v_\perp}{r} \frac{\partial f}{\partial v_\perp} = 0 , \qquad (12)$$

$$\frac{1}{r} \frac{d^2(r\psi)}{dr^2} = \pi \int_0^\infty d(v_\perp^2) \int_{-\infty}^\infty dv_r f - 1 . \qquad (13)$$

In Ref. 31, we have given a simple, exact solution of (12) and (13) just to illustrate the essential physics of such spherically symmetric solutions. As shown in Fig. 4, the basic idea in constructing such solutions is to place electrons in circular orbits around a spherically symmetric electrostatic potential in such a way that the density of electrons decreases towards the central region. This ensures that the net charge density is positive around the center, which, in turn, produces self-consistently the required electric potential needed for bounded solutions. Many possible solutions $\psi(r)$ can be constructed in this way, for example, $\psi(r) = \psi_0 \exp(-r^2/\delta^2)$ with the condition $\delta^2 > 6\psi_0$, which is a width-amplitude relation of the electrostatic potential.

For a more detailed discussion of other such width-amplitude inequalities that are relevant to observations in space, the reader is referred to Ref. 58.

The simple example discussed above shows that if we can arrange electrons to circulate around the potential and to allow dependence on angular momentum, it becomes possible to construct 3D solutions. We now consider a more general solution of (12), $f = f(w,l)$, where $l = v_\perp r$ is the angular momentum. The next step is to find specific solutions of this form that also satisfy (13). For simplicity, we consider solutions of the separable form $f(w,l) = (2\pi)^{-3/2} \exp(-w) f_1(l)$. Since we seek a solution localized in $r$, we impose the boundary condition that as $\psi \to 0$, the distribution function must tend to a Maxwellian, that is, $f \to (2\pi)^{-3/2} \exp(-w)$. Hence, $f_1 \to 1$ as $r \to \infty$. For specificity, we choose

$$f_1(v_\perp r) = 1 - (1 - h_0) \exp(-v_\perp^2 r^2 / x_0^2) , \qquad (14)$$

with two real parameters $h_0$ and $x_0$. We note that $f_1(0) = h_0 \geq 0$, and $f_1(\infty) = 1$. Eq. (13) then becomes

$$\frac{1}{r} \frac{d^2(r\psi)}{dr^2} = e^\psi h(r) - 1 , \qquad (15)$$

where we obtain by direct integration, $h(r) = (h_0 + 2r^2/x_0^2)/(1 + 2r^2/x_0^2)$. Equation (15) is a second-order non-linear ordinary differential equation for $\psi(r)$, to be solved subject to the boundary conditions $\psi(r \to \infty) \to 0$, $\psi(r = 0) = \psi_0$, and $\psi'(r = 0) = 0$. For the special case $h_0 = 1$, it follows that $f_1 = h = 1$ for all $r$. Then the distribution function depends only on energy. In this case, by our discussion above, there is no non-trivial solution. That this is so can also be seen from (15), which yields only the trivial solution $\psi(r) \equiv 0$ if $h = 1$. If this were not so, then for any non-zero positive (negative) value of $\psi_0$, $\psi(r \to \infty)$ will tend to positive (negative) infinity, which violates the boundary condition as $r \to \infty$. Similarly[31], it can be shown that when $h_0 \neq 1$, a well-behaved solution must exist for a certain $\psi_0$.

Figure 5(a) shows the solution $\psi(r)$ for the case with $h_0 = 0.9$ and $x_0 = 1$. The same solution is plotted on log-log scale in Figure 5(b) showing the asymptotic behavior of $\psi(r \to \infty) \to \psi_\infty / r^2$ for a constant $\psi_\infty$. In fact, by directly substituting this asymptotic form into (15), the value of $\psi_\infty$ is found to be $\psi_\infty = x_0^2 (1 - h_0)/2$. This also means that the electric field has the asymptotic behavior $\mathbf{E}(r \to \infty) \to 2\psi_\infty \hat{\mathbf{r}}/r^3$. Since the electric field falls off faster than $1/r^2$, the global solution is asymptotically charge-neutral. We plot the radial electric field in Figure 5(c), and the normalized charge density $\rho_q(r) \equiv 1 - e^\psi h(r)$ in Figure 5(d). Careful inspection shows a long tail for $\rho_q$ moving from positive to negative values at $r \sim 2.7$.

For the case $h_0 > 1$, we can demonstrate similarly that there must exist a negative $\psi_0$ that satisfies the required boundary condition $\psi(r \to \infty) \to 0$. Figure 6 shows the solutions when $h_0 = 1.1$ and $x_0 = 1$, which are very similar to the solutions represented in Figure 5 except that their signs are reversed. This is an interesting case because a negative potential is supported in 3D entirely by electron dynamics. It would be impossible to realize such a solution in 1D since all electrons in such a potential will have passing trajectories that will tend to infinity, where the distribution must be a Maxwellian. This will not support a localized BGK solution with negative potential. The only way a localized BGK solution with negative potential can be supported in 1D is by taking into account the dynamics of trapped ions.

It is interesting to determine the relation between the potential $\psi_0$ and the characteristic length $x_0$. Figure 7 shows the plots of $\psi_0$ vs. $x_0$ for various values of $h_0$. For large $x_0$, $\psi_0$ simply tends to $\ln(1/h_0)$. As $x_0 \to 0$, we see that $\psi_0 \to 0$. In this limit, the functional dependence is very similar to that of $\psi_\infty$, except that it has a larger numerical coefficient, that is, $\psi_0 \sim 8x_0^2(1-h_0)$. We note that even in this limit, the solution itself as well as the potential $\psi$ cannot be recovered by solving the linearized Vlasov equation since the velocity gradient of $f$ is large around $v_\perp \to 0$. In other words, this is a truly nonlinear solution even in the small amplitude limit.

We remark that the width-amplitude inequality based on these solutions is very similar to that given in Ref. 58, which have been shown to be consistent with observations. The feature that the electrostatic potential of these 3D solutions is typically single-humped in all directions is consistent with observations that solitary waves in the auroral ionosphere are bipolar in the parallel electric field $E_\parallel$ and unipolar in both components of perpendicular electric field $E_\perp$. It is interesting that these features are realized even without including the background magnetic field in our solutions.

We should emphasize that the solutions presented here are a few among many, chosen to illustrate the principles of our approach. One major extension of our approach will be the inclusion of possible piecewise distribution functions distinguishing trapped and passing electrons. Such distinction is not made in the solutions presented here, nor are they necessary because we are able to describe the whole distribution function with a single continuous and smooth function. This should be contrasted with the 1D case, where the distinction between trapped and passing particles must be made in order to construct a solution.

## IV. 2D BGK Modes with Finite Magnetic Field

Although the 3D solutions found in the last section pertain to unmagnetized plasmas, the technique developed there can be applied to study solutions with a finite magnetic field. As mentioned above, 3D solutions obtained so far for the magnetized case are those with a strong magnetic field such that one can apply the guiding-center approximation. As such, they are not exact steady-state solutions for the Vlasov-Poisson equations. In order to see if these solutions can survive when finite-Larmor-radius effects are taken into account, it is important to see if exact 3D BGK modes exist with a finite magnetic field. First, for the cylindrically symmetric case with $\psi = \psi(\rho,z)$, $f = f(\rho,z,v_\rho,v_\phi,v_z)$ and a uniform background magnetic field $\mathbf{B} = B\hat{\mathbf{z}}$, the Vlasov equation takes the form

$$v_\rho \frac{\partial f}{\partial \rho} + v_z \frac{\partial f}{\partial z} + \left( \frac{\partial \psi}{\partial \rho} + \frac{v_\phi^2}{\rho} - Bv_\phi \right) \frac{\partial f}{\partial v_\rho} - \left( \frac{v_\rho v_\phi}{\rho} - Bv_\rho \right) \frac{\partial f}{\partial v_\phi} + \frac{\partial \psi}{\partial z} \frac{\partial f}{\partial v_z} = 0. \tag{16}$$

The Vlasov equation can then be solved by a distribution function of the form $f = f(w,l)$, where $l = 2\rho v_\phi - B\rho^2$ is proportional to the angular momentum. To find a BGK solution, one then needs to solve the Poisson equation

$$\frac{1}{\rho} \frac{\partial}{\partial \rho} \left( \rho \frac{\partial \psi}{\partial \rho} \right) + \frac{\partial^2 \psi}{\partial z^2} = \int d^3v f\left( \frac{v^2}{2} - \psi, 2\rho v_\phi - B\rho^2 \right) - 1, \tag{17}$$

self-consistently. The result is a partial differential equation for $\psi$ in two variables instead of the ordinary differential equation (15), and thus substantially more analysis is needed to discuss its solution. Therefore, we will leave the full 3D case for later study and consider here only the special 2D case with $\psi = \psi(\rho)$. Due to the same reason as given in the last section, it is impossible to have a localized BGK solution with $f = f(w)$ only. Instead, we consider a solution of the form

$$f(w,l) = (2\pi)^{-3/2} \exp(-w)\left[1 - h_0 \exp(-kl^2)\right], \tag{18}$$

where the two constant parameters satisfy $1 > h_0 > 0$, and $k > 0$. Performing the integration in (17) gives

$$\frac{1}{\rho}\frac{d}{d\rho}\left(\rho\frac{d\psi}{d\rho}\right) = e^{\psi(\rho)}\left[1 - \frac{h_0}{\sqrt{1+8k\rho^2}}\exp\left(-\frac{kB^2\rho^4}{1+8k\rho^2}\right)\right] - 1. \tag{19}$$

One then needs to solve this equation for $\psi(\rho)$, subject to boundary conditions $\psi(\rho \to \infty) \to 0$, $\psi(\rho = 0) = \psi_0$, and $\psi'(\rho = 0) = 0$. Using an argument similar to that given in the discussion of (15), we see that there should be a solution for a certain $\psi_0$. Equation (19) can then be solved numerically. Figure 8(a) shows the solution $\psi(\rho)$ for the case with $h_0 = 0.1$, $k = 1$ and $B = 1$. The same solution is plotted on semi-log scale in Figure 8(b) showing the asymptotic behavior, $\psi(\rho \to \infty) \propto \exp(-\rho)$. This also means that the electric field vanishes asymptotically and thus the global solution is asymptotically charge-neutral. We plot the radial electric field in Figure 8(c), and the normalized charge density $\rho_q(\rho)$ in Figure 8(d), which also moves from positive to negative at some point in $\rho$. Overall, this solution is very similar in shape to the 3D solutions in the last section, except for the asymptotic behavior. Again, the above solution is just one of many possible classes of 2D BGK modes.

There is, however, one fundamental difference between these 2D solutions and the usual 1D BGK modes (or equivalently, the 3D BGK modes with infinite magnetic field[29-30], and the 3D BGK modes without magnetic field): these solutions carry finite current density. In principle, such current density should produce a magnetic field through Ampère's law,

$$\nabla \times \mathbf{B} = -\beta_e^2 \int d^3v f \mathbf{v}, \tag{20}$$

where the equation is normalized as before, and $\beta_e = v_e/c$ is the ratio between electron thermal speed and the speed of light. In order to complete the construction of these solutions, we, need to test that the constructed solutions satisfy, not just the Vlasov-Poisson system, but the more complete electromagnetic Vlasov-Poisson-Ampère system. For the case with $\beta_e \ll 1$, one would expect the electromagnetic correction to the Vlasov-Poisson solutions to be small. It turns out that, for this special 2D case, it is straightforward to implement such a correction exactly. First, one recognizes that (18), the Vlasov equation with magnetic field, can be generalized to the case with a cylindrically symmetric magnetic field $\mathbf{B} = \nabla \times [A_\phi(\rho,z)\hat{\phi}]$. The solution will still have the general form of $f = f(w,l)$, except now $l = 2\rho(v_\phi - A_\phi)$. Note that for a uniform magnetic field $\mathbf{B} = B_0\hat{\mathbf{z}}$, we obtain $A_\phi = B_0\rho/2$. The form of the above solutions will remain the same, except that we need to replace $B$ with $2A_\phi/\rho$ in (19), with $A_\phi = A_\phi(\rho)$ for the 2D case. With this solution, the integration in (20) can be performed to obtain the equation,

$$\frac{d}{d\rho}\left(\frac{1}{\rho}\frac{d(\rho A_\phi)}{d\rho}\right) = -\frac{8\beta_e^2 h_0 k\rho^2 A_\phi}{(1+8k\rho^2)^{3/2}}\exp\left(\psi - \frac{4kA_\phi^2\rho^2}{1+8k\rho^2}\right). \tag{21}$$

The two equations (19) and (21) have to be solved together to obtain a self-consistent solution for both $\psi$ and $A_\phi$. The boundary conditions for (21) are $A_\phi = B_0\rho/2 + A_{\phi 1}$ with $A_{\phi 1}(0) = 0$ and $A_{\phi 1}(\rho \to \infty) \to 0$ so that we have a uniform magnetic field $\mathbf{B} = B_0\hat{\mathbf{z}}$ asymptotically. Since we have a small parameter in $\beta_e$, we will solve these equations iteratively. First, we use $A_\phi = B_0\rho/2$ for a uniform magnetic field to calculate $\psi$. We then substitute $\psi$ and $A_\phi$ into the right hand side of (21) to solve for a new $A_\phi$. The process is then repeated until both $\psi$ and $A_\phi$ converges. When converged, this is an exact solution of the steady-state Vlasov-Poisson-Ampère (or Vlasov-Maxwell) equations, instead of just the Vlasov-Poisson equations. We have implemented this scheme numerically. The convergence is indeed very fast for small $\beta_e$, and the correction to the original BGK mode is also small, as expected. However, we do find that for some choice of $h_0$, $k$, and $B_0$, significant corrections are needed, after a longer process of iteration for weakly relativistic case with $\beta_e \sim 0.1$. Of course, our non-relativistic treatment in this paper will have to be generalized for even higher electron thermal speed. Due to limitation of space, we postpone a more detailed discussion of these solutions to a separate publication.

**IV. Conclusions**

We have presented our recent results on the classical problems of Landau damping and BGK modes. In the problem of Landau damping, we have shown that even a weak collision can have profound implications for the theory, and changes completely the nature of the spectrum. We have shown that the Case-Van Kampen continuum modes, which comprise a complete spectrum for the collisionless problem, are eliminated and replaced by a complete set of eigenfunctions, which is totally different from the Case-Van Kampen continuum in the collisionless case.

In the problem of BGK modes, we have shown that a 3D exact solution can be constructed in an unmagnetized plasma if and only if the distribution function is not only dependent on energy, but also on other constants of motion such as the angular momentum. We have constructed a class of such solutions, analyzed their functional form, and their width-amplitude inequalities. For a plasma with a finite magnetic field, we have constructed exact 2D BGK modes for the Vlasov-Poisson-Ampère equations. It remains to be shown if 3D BGK modes for a finite magnetic field can exist. Much work remains to be done beyond what is presented here, but we hope that the solutions presented here can provide guidance and insight.

As shown in the shape of the eigenfunctions for the Landau damping problem with weak collision, sharp gradients develop generically in velocity space. Therefore, nonlinear effects can be very important. Such sharp gradients can also exist in BGK modes. It will be interesting to investigate how weak collisions can alter the results of BGK theory. The combined effect of nonlinearity and weak collision is left to future investigations.

As mentioned in the introduction, the mathematical framework of the Vlasov-Poisson system can be applied to many different areas of physics. Our results here may provide new insights in related problems in these areas. Conversely, some of the results presented here can also be anticipated by progress made in other areas. For example, in the problem of vorticity defects in viscous shear, a formulation similar to the weakly collisional Landau damping problem has been developed[68]. Some eigenmodes in that problem in the small-viscosity limit are found to reduce to the so-called quasi-modes, similar to the Landau damped eigenmodes in our treatment (see Refs. 68, 69 and references therein). However, it appears that the completeness of these eigenmodes was not discussed in the cited papers. It may be possible to adapt our proof of completeness to the problem of vorticity defects. This is left to future work.

This research is supported by the Department of Energy and the Paul Chair Endowment at the University of New Hampshire.

Figure Captions

Figure 1. Linear amplitude contours of the transformed ion response function, plotted in terms of the particle velocity and of the phase velocity. The phase velocities calculated from kinetic eigenmodes are indicated for comparison.

Figure 2. Eigenvalues on the complex $\kappa$ plane for the spatial problem with $\alpha = 1.6$. ($\diamond$) $\mu = 0.1$; ($\ast$) $\mu = 0.05$; (+) $\mu = 0.025$ (×) Landau roots ($\mu = 0$).

Figure 3. Solid curves: eigenfunctions corresponding to the least-damped eigenvalue $\Omega_0$ for the temporal problem with $\alpha = 9$. Dashed curves: the function $\eta(u)/(u - \Omega_0)$. (a) and (b) Real and imaginary parts of the eigenfunction, respectively, for $\mu = 0.025$. (c) and (d) Real and imaginary parts of the eigenfunction, respectively, for $\mu = 0.000391$.

Figure 4. Schematic illustration of the role of electron angular momentum in realizing 3D BGK solutions. The radial electric field provides the centripetal force necessary to sustain the circular electron motion.

Figure 5. (a) Numerical solution $\psi(r)$ for the case with $h_0 = 0.9$ and $x_0 = 1$. (b) The same solution in log-log plot. The dashed line shows the relation $\psi(r \to \infty) \to \psi_\infty / r^2$. (c) Radial electric field. (d) Normalized charge density $\rho_q(r) \equiv 1 - e^\psi h(r)$.

Figure 6. Same as in Figure 5 for the case $h_0 = 1.1$ and $x_0 = 1$. The profiles are nearly identical as those in Figure 5, except that the signs are reversed.

Figure 7. Plots of $\psi_0$ versus $x_0$ for various values of $h_0$, from 0.1 (top curve) to 0.9 (bottom curve) in an increment of 0.1. The dashed line is $\psi_0 = 8x_0^2(1-h_0)$ for $h_0 = 0.1$

Figure 8. (a) Numerical solution $\psi(\rho)$ for the case with $h_0 = 0.1$, $k = 1$ and $B = 1$. (b) The same solution in semi-log plot. The dashed line is $\psi = 0.1\exp(-\rho)$. (c) Radial electric field. (d) Normalized charge density $\rho_q(\rho)$.

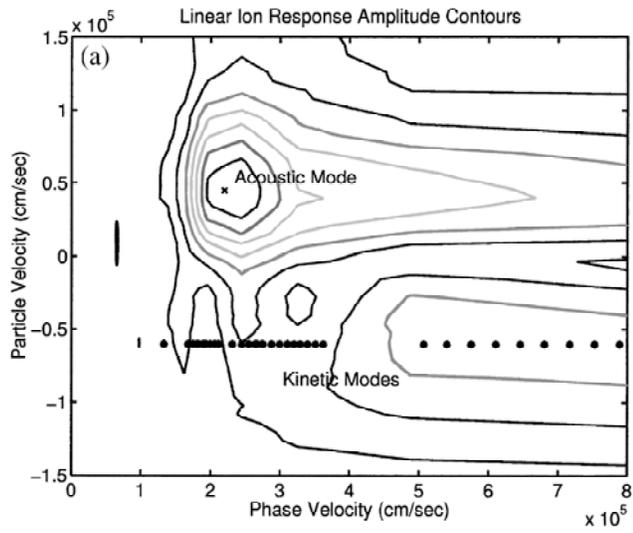

Figure 1

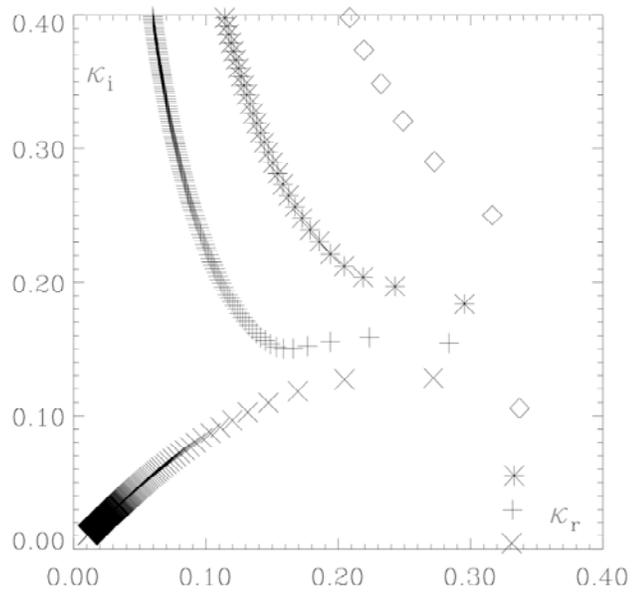

Figure 2

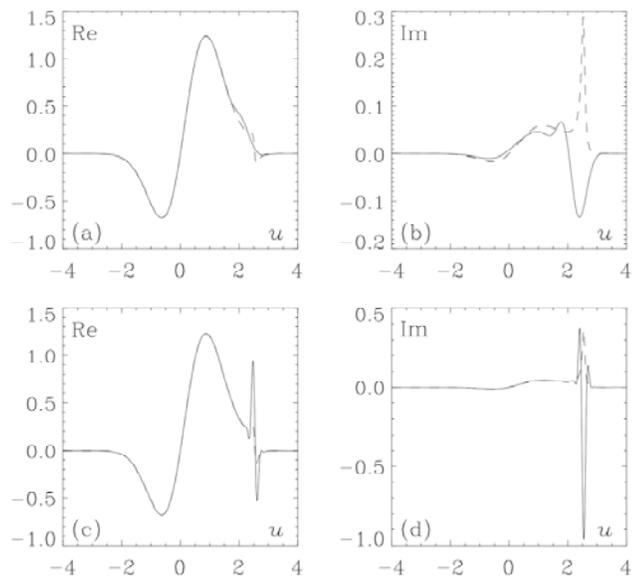

Figure 3

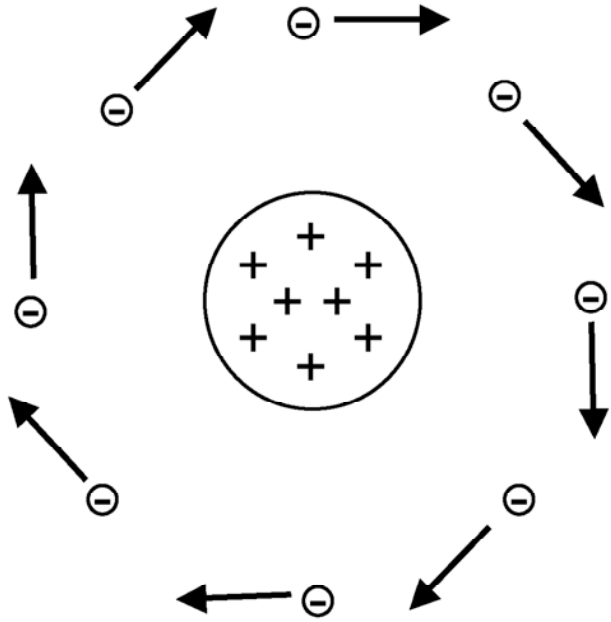

Figure 4

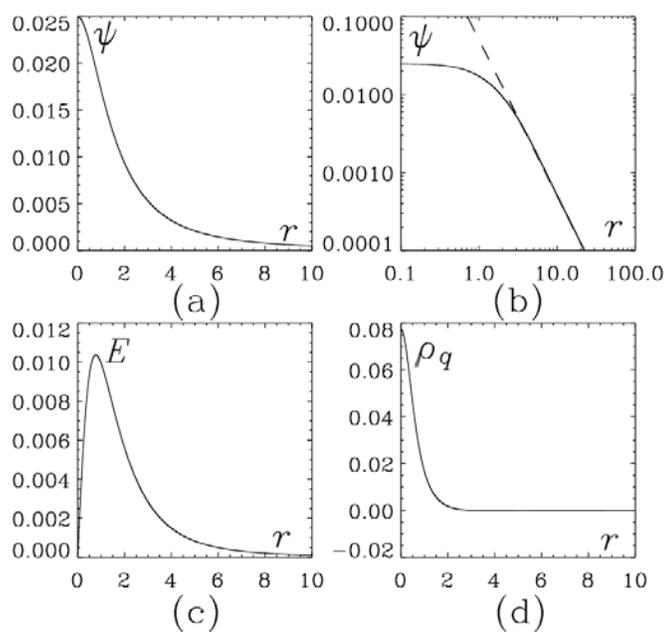

Figure 5

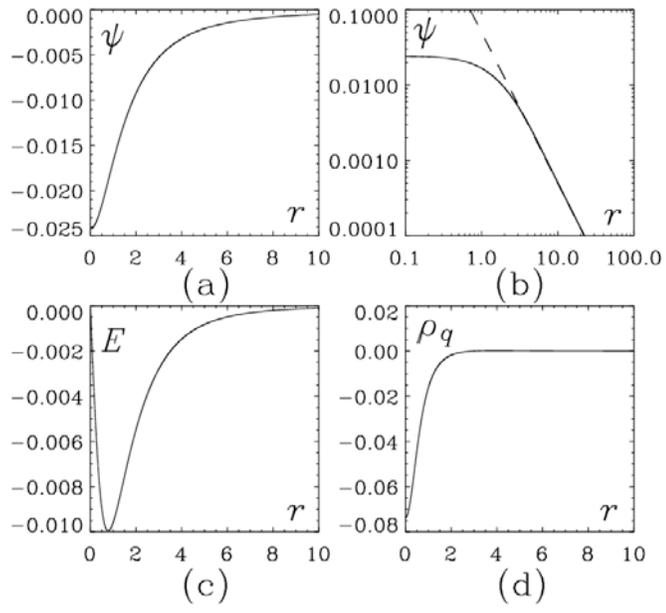

Figure 6

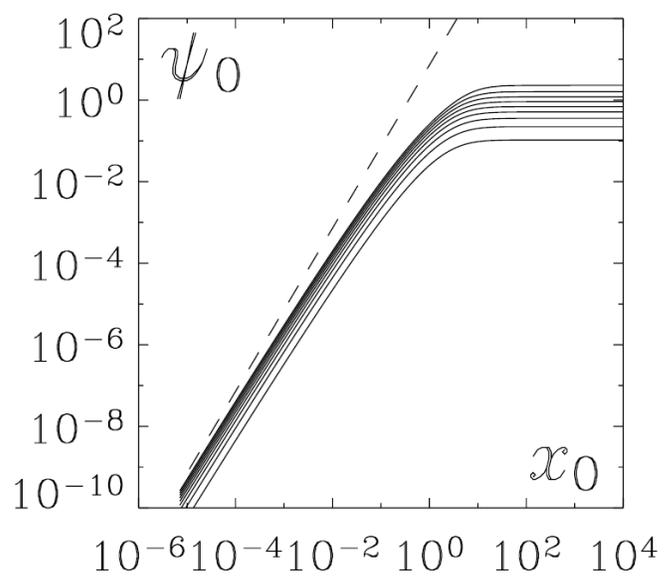

Figure 7

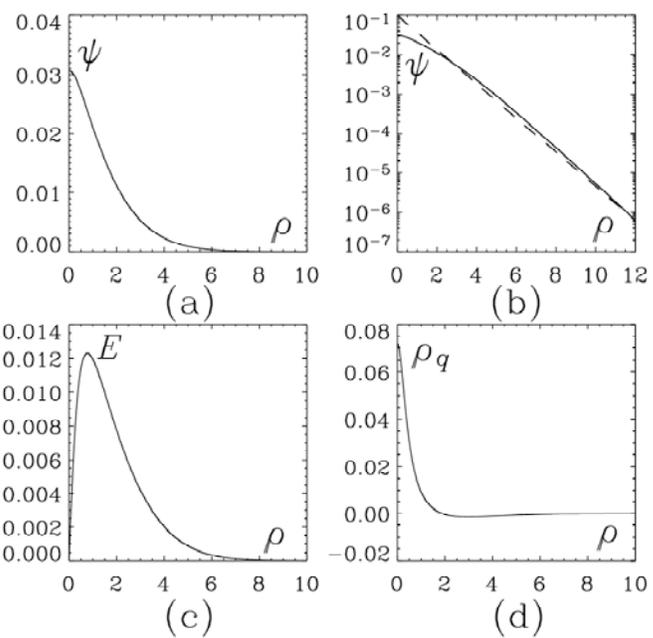

Figure 8